%% file: main.tex
\documentclass[
  sigconf
]{acmart}

\usepackage{booktabs} %
\usepackage{balance}
\usepackage[utf8]{inputenc}
\usepackage{tikz}
\usetikzlibrary{automata,positioning}

\usepackage{subcaption}
\usepackage[ruled]{algorithm2e} %
\usepackage{url}

\setcopyright{rightsretained}
\acmDOI{}

\acmISBN{}

\acmConference[arxiv]{arxiv}{May 03, 2019}{Munich}
\acmYear{2019}
\copyrightyear{2019}

\begin{document}
\title[On the Impact of Memory Allocation on High-Performance Query Processing]{On the Impact of Memory Allocation\\ on High-Performance Query Processing}

\author{Dominik Durner}
\affiliation{%
  \institution{Technische Universität München}
}
\email{dominik.durner@tum.de}

\author{Viktor Leis}
\affiliation{%
  \institution{Friedrich-Schiller-Universität Jena}
}
\email{viktor.leis@uni-jena.de}

\author{Thomas Neumann}
\affiliation{%
  \institution{Technische Universität München}
}
\email{thomas.neumann@tum.de}

\begin{abstract}
Somewhat surprisingly, the behavior of analytical query engines is crucially affected by the dynamic memory allocator used.
Memory allocators highly influence performance, scalability, memory efficiency and memory fairness to other processes.
In this work, we provide the first comprehensive experimental analysis on the impact of memory allocation for high-performance query engines.
We test five state-of-the-art dynamic memory allocators and discuss their strengths and weaknesses within our DBMS.
The right allocator can increase the performance of TPC-DS (SF 100) by $2.7$x on a 4-socket Intel Xeon server.
\end{abstract}
\maketitle

\input{chapters/introduction}

\input{chapters/related}
\input{chapters/malloc}
\input{chapters/workload}

\input{chapters/analysis}

\input{chapters/conclusion}

\balance
\bibliographystyle{acmref}
\bibliography{bibliography}

\end{document}

%% file: chapters/introduction.tex
\section{Introduction}

Modern high-performance query engines are orders of magnitude faster than traditional database systems.
As a result, components that hitherto were not crucial for performance may become a performance bottleneck.
One such component is memory allocation.
Most modern query engines are highly parallel and heavily rely on temporary hash-tables for query processing which results in a large number of short living memory allocations of varying size.
Memory allocators therefore need to be scalable and be able to handle myriads of small and medium sized allocations as well as several huge allocations simultaneously.
As we show in this paper, memory allocation has become a large factor in overall query processing performance.

New hardware trends exacerbate the allocation issues.
The development of multi- and many-core server architectures with up to hundred general purpose cores is a distinct challenge for memory allocation strategies.
Due to the increased number of pure computation power, more active queries are possible.
Furthermore, multi-threaded data structure implementations lead to dense and simultaneous access patterns.
Because most multi-node machines rely on a non-uniform memory access (NUMA) model, requesting memory from a remote node is particularly expensive.

Therefore, the following goals should be accomplished by a dynamic memory allocator:
\begin{description}
  \item[Scalability] Reduce overhead for multi-threaded allocations.
  \item[Performance] Minimize the overhead for malloc and free.
  \item[Memory Fairness] Give freed memory back to the OS.
  \item[Memory Efficiency] Avoid memory fragmentation.
\end{description}

\begin{figure}[tb]
  \centering
  \includegraphics{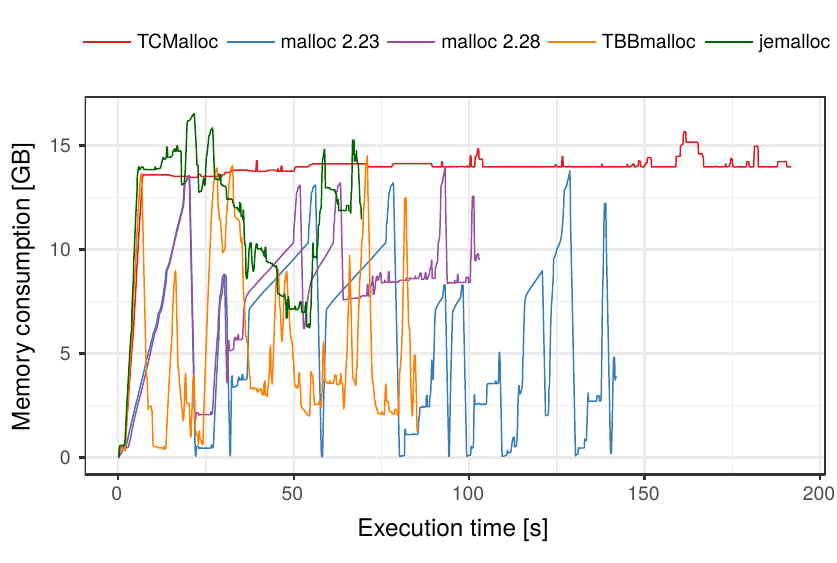}
  \vspace{-1em}
  \caption{Execution of a given query set on TPC-DS (SF 100) with different allocators.}
  \label{fig:mem_1_10}
  \vspace{-1em}
\end{figure}

In this paper, we perform the first comprehensive study of memory allocation in modern database systems.
We evaluate different approaches to the aforementioned dynamic memory allocator requirements.
Although memory allocation is on the critical path of query processing, no empirical study on different dynamic memory allocators for in-memory database systems has been conducted~\cite{DBLP:journals/pvldb/AppuswamyAPIA17}.

Figure~\ref{fig:mem_1_10} shows the effects of different allocation strategies on TPC-DS with scale factor 100.
We measure memory consumption and execution time with our multi-threaded database system on a 4-socket Intel Xeon server.
In this experiment, our DBMS executes the query set sequentially using all available cores.
Even this relatively simple workload already results in significant performance and memory usage differences.
Our database linked with \texttt{jemalloc} can reduce the execution time to $\frac{1}{2}$ in comparison to linking it with the standard \texttt{malloc} of glibc 2.23.
On the other hand, \texttt{jemalloc} has the highest memory consumption and does not release memory directly after execution of the query.
Although the resident memory consumption seems high for \texttt{TCMalloc}, it already gives back the memory to the operating system lazily.
Consequently, the allocation strategy is crucial to the performance and memory consumption behavior of in-memory database systems.

The rest of this paper is structured as follows:
After discussing related work in Section~\ref{sec:related}, we describe the used allocators and their most important design details in Section~\ref{sec:malloc}.
Section~\ref{sec:workload} highlights important properties of our DBMS and analyzes the executed workload according to its allocation pattern.
Our comprehensive experimental study is evaluated in Section~\ref{sec:study}.
Section~\ref{sec:summary} summarizes our findings.

%% file: chapters/related.tex
\section{Related Work}
\label{sec:related}

Although memory allocation is a major performance driver, no empirical study on the impact on in-memory database systems has been conducted.
Ferreira et al.~\cite{ferreira2011experimental} analyzed dynamic memory allocators for a variety of multi-threaded workloads.
However, the study considers only up to 4 cores.
Therefore, it is hard to predict the scalability for today's many-core systems.

In-memory DBMS and analytical query processing engines, such as HyPer~\cite{DBLP:conf/icde/KemperN11}, SAP HANA~\cite{DBLP:conf/btw/May0L17}, and Quickstep~\cite{DBLP:journals/pvldb/PatelDZPZSMS18} are built to utilize as many cores as possible to speed up query processing.
Because these system rely on allocation-heavy operators (e.g., hash joins, aggregations), a revised experimental analysis on the scalability of the state-of-the-art allocators is needed.
In-memory hash joins and aggregations can be implemented in many different ways which can influence the allocation pattern heavily~\cite{DBLP:conf/icde/BalkesenTAO13,DBLP:conf/sigmod/BlanasLP11, DBLP:conf/cidr/ZhangDP19, DBLP:conf/sigmod/LeisBK014}.

Some online transaction processing (OLTP) systems try to reduce the allocation overhead by managing their allocated memory in chunks to increase performance for small transactional queries~\cite{DBLP:conf/sosp/TuZKLM13, DBLP:conf/damon/StoicaA13, durner}.
However, most database systems process both transactional and analytical queries.
Therefore, the wide variety of memory allocation patterns for analytical queries needs to be considered as well.
Custom chunk memory managers help to reduce memory calls for small allocations but larger chunk sizes trade memory efficiency in favor of performance.
Thus, our database system uses transaction-local chunks to speed up small allocations.
Despite these optimizations, allocations are still a performance issue.
Hence, the allocator choice is crucial to maximize throughput.

With the development of non-volatile memory (NVM), new allocation requirements were introduced.
Foremost, the defragmentation and safe release of unused memory is important since all changes are persistent.
New dynamic memory allocators for these novel persistent memory systems have been developed and experimentally studied~\cite{DBLP:journals/pvldb/OukidBLLWG17}.
However, regular allocators outperform these NVM allocators in most workloads due to fewer memory constraints.

%% file: chapters/malloc.tex
\section{Memory Allocators}
\label{sec:malloc}

In this section, we discuss the five different allocation strategies used for our experimental study. %
We explain the basic properties of these algorithms according to memory allocation and freeing.
The tested state-of-the-art allocators are available as Ubuntu~18.10 packages.
Only the glibc \texttt{malloc 2.23} implementation is part of a previous Ubuntu package.
Nevertheless, this version is still used in many current distributions such as the stable Debian release.

Memory allocation is strongly connected with the operating system (OS).
The mapping between physical and virtual memory is handled by the kernel.
Allocators need to request virtual memory from the OS.
Traditionally, the user program asks for memory by calling the malloc method of the allocator.
The allocator either has memory available that is unused and suitable or needs to request new memory from the OS.
For example, the Linux kernel has multiple APIs for requesting and freeing memory.
\texttt{brk} calls can increase and decrease the amount of memory allocated to the data segment by changing the program break.
\texttt{mmap} maps files into memory and implements demand paging such that physical pages are only allocated if used.
With anonymous mappings, virtual memory that is not backed by a real file can be allocated within main memory as well.
The memory allocation process is visualized below.

\begin{center}
\vspace{0.25em}
\includegraphics[width=\columnwidth]{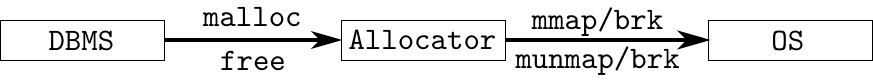}
\end{center}

Besides freeing memory directly with the aforementioned calls, the memory allocator can opt to release memory with \texttt{MADV\_FREE} (since Linux Kernel 4.5).
\texttt{MADV\_FREE} indicates that the kernel is allowed to reuse this memory region.
However, the allocator can still access the virtual memory address and either receives the previous physical pages or the kernel provides new zeroed pages.
Only if the kernel reassigns the physical pages, new ones need to be zeroed.
Hence, \texttt{MADV\_FREE} reduces the number of pages that require zeroing since the old pages might be reused by the same process.

\subsection{\texttt{malloc 2.23}}
The glibc \texttt{malloc} implementation is derived from \texttt{ptmalloc2} which originated from \texttt{dlmalloc}~\cite{mallocdoc}. %
It uses chunks of various sizes that exist within a larger memory region known as the heap.
\texttt{malloc} uses multiple heaps that grow within their address space.

For handling multi-threaded applications, \texttt{malloc} uses arenas that consist of multiple heaps.
At program start the main arena is created and additional arenas are chained with previous arena pointers.
The arena management is stored within the main heap of that arena.
Additional arenas are created with \texttt{mmap} and are limited to eight times the number of CPU cores.
For every allocation, an arena-wide mutex needs to be acquired.
Within arenas free chunks are tracked with free-lists.
Only if the top chunk (adjacent unmapped memory) is large enough, memory will be returned to the OS.

\vspace{0.15em}
\begin{center}
\includegraphics[scale=0.8]{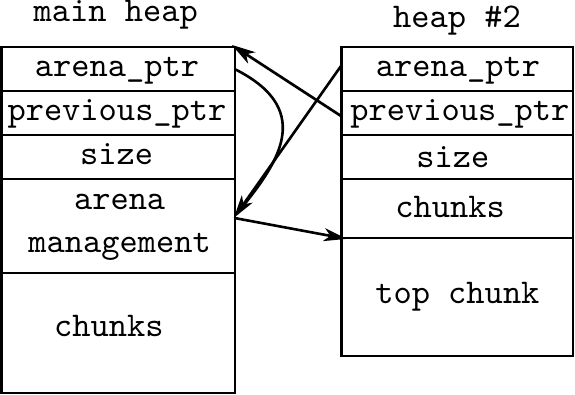}
\end{center}
\vspace{0.15em}

\texttt{malloc} is aware of multiple threads but no further multi-threaded optimizations, such as thread locality or NUMA awareness, is integrated.
It assumes that the kernel handles these issues.

\subsection{\texttt{malloc 2.28}}
A thread-local cache (tcache) was introduced with glibc v2.26~\cite{malloc226}.
This cache requires no locks and is therefore a fast path to allocate and free memory.
If there is a suitable chunk in the tcache for allocation, it is directly returned to the caller bypassing the rest of the malloc routine.
The deletion of a chunk works similarly.
If the tcache has a free slot, the chunk is stored within it instead of immediately freeing it.

\subsection{\texttt{jemalloc 5.1}}
\texttt{jemalloc} was originally developed as scalable and low fragmentation standard allocator for FreeBSD.
Today, \texttt{jemalloc} is used for a variety of applications such as Facebook, Cassandra and Android.
It differentiates between three size categories - small ($< 16/text{KB}$), large ($< 4\text{MB}$) and huge.
These categories are further split into different size classes.
It uses arenas that act as completely independent allocators.
Arenas consist of chunks that allocate multiples of 1024 pages ($4\text{MB}$).
\texttt{jemalloc} implements low address reusage for large allocations to reduce fragmentation.
Low address reusage, which basically scans for the first large enough free memory region, has similar theoretical properties as more expensive strategies such as best-fit.
\texttt{jemalloc} tries to reduce zeroing of pages by deallocating pages with \texttt{MADV\_FREE} instead of unmapping them.
Most importantly, \texttt{jemalloc} purges dirty pages decay-based with a wall-clock (since v4.1) which leads to a high reusage of recently used dirty pages.
Consequently, the unused memory will be purged if not requested anymore to achieve memory fairness~\cite{Evans:2015:TTM:2742580.2742807, jemallocchangelog}.

\vspace{0.15em}
\begin{center}
\includegraphics[scale=0.8]{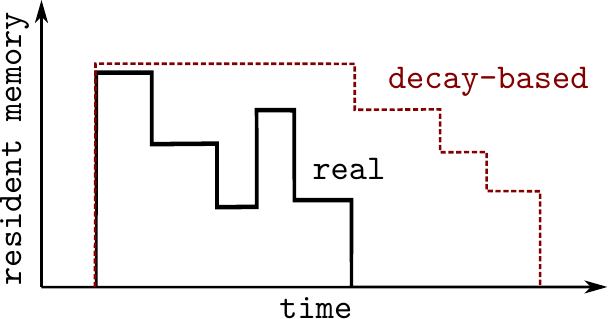}
\end{center}

\subsection{\texttt{TBBmalloc 2017 U7}}
Intel's Threading Building Blocks (TBB) allocator is based on the scalable memory allocator \texttt{McRT}~\cite{DBLP:conf/iwmm/HudsonSAH06}.
It differentiates between small, quite large, and huge objects.
Huge objects ($\ge 4\text{MB}$) are directly allocated and freed from the OS.
Small and large objects are organized in thread-local heaps with chunks stored in memory blocks.

Memory blocks are memory mapped regions that are multiples of the requested object size class and inserted into the global heap of free blocks.
Freed memory blocks are stored within a global heap of abandoned blocks.
If a thread-local heap needs additional memory blocks, it requests the memory from one of the global heaps.
Memory regions are unmapped during coalescing of freed memory allocations if no block of the region is used anymore~\cite{kukanov2007foundations, tbbmalloc}.

\subsection{\texttt{TCMalloc 2.5}}
\texttt{TCMalloc} is part of Google's gperftools.
Each thread has a local cache that is used to satisfy small allocations ($\le 256\text{KB}$).
Large objects are allocated in a central heap using $8\text{KB}$ pages.

\texttt{TCMalloc} uses different allocatable size classes for the small objects and stores the thread cache as a singly linked list for each of the size classes.
Medium sized allocations ($\le 1\text{MB}$) use multiple pages and are handled by the central heap.
If no space is available, the medium sized allocation is treated as a large allocation.
For large allocations, spans of free memory pages are tracked within a red-black tree.
A new allocation just searches the tree for the smallest fitting span.
If no span is found the memory is allocated from the kernel~\cite{tcmallocdoc}.

Unused memory is freed with the help of \texttt{MADV\_FREE} calls.
Small allocations will be garbage collected if the thread-local cache exceeds a maximum size.
Freed spans are immediately released since the aggressive decommit option was enabled (starting with version 2.3) to reduce memory fragmentation~\cite{tcmallocrepo}.

%% file: chapters/workload.tex
\section{DBMS and Workload Analysis}
\label{sec:workload}

Decision support systems rely on analytical queries (OLAP) that gather information from a huge dataset by joining different relations for example.
In in-memory query engines joins are often scheduled physically as hash joins resulting in a huge number of smaller allocations.
In the following, we use a database system that uses pre-aggregation hash tables to perform multi-threaded group bys and joins~\cite{DBLP:conf/sigmod/LeisBK014}.
Our DBMS has a custom transaction-local chunk allocator to speed up small allocations of less than $32\text{KB}$.
We store small allocations in chunks of medium sized memory blocks.
Since only small allocations are stored within chunks, the memory efficiency footprint of these small object chunks is marginal.
Additionally, the memory needed for tuple materialization is acquired in chunks.
These chunks grow as more tuples are materialized.
Thus, we already reduce the stress on the allocator significantly while preserving memory efficiency.

The TPC-H and TPC-DS benchmarks were developed to standardize common decision support workloads~\cite{DBLP:conf/vldb/OthayothP06}.
Because TPC-DS contains a larger workload of more complex queries than TPC-H, we focus on TPC-DS in the following.
As a result, we expect to see a more diverse and challenging allocation pattern.
TPC-DS describes a retail product supplier with different sales channels such as stores and web sales.

\begin{figure}
  \centering
  \begin{subfigure}[t]{0.25\textwidth}
    \includegraphics{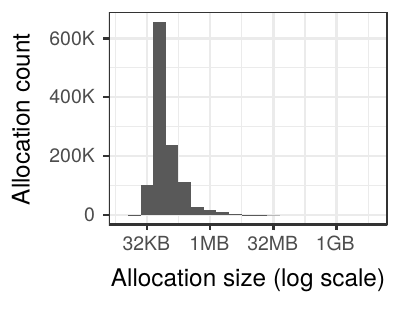}
    \caption{By number of allocations.}
  \end{subfigure}%
  \begin{subfigure}[t]{0.25\textwidth}
    \includegraphics{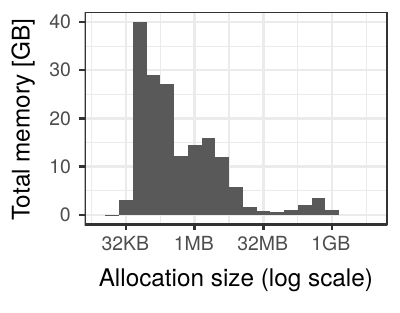}
    \caption{By memory consumption.}
  \end{subfigure}

  \caption{Allocations in TPC-DS (SF 100, serial execution).}
  \label{fig:allocpattern}
  \vspace{-1em}
\end{figure}

In the following, we statistically analyze the allocation pattern for TPC-DS executing all queries without rollup and window functions.
Note that the specific allocation pattern depends on the discussed implementation choices of the join and group by operators.

Figure~\ref{fig:allocpattern} shows the distribution of allocations in our system for TPC-DS with scale factor 100.
The most frequent allocations are in the range of $32\text{KB}$ to $512\text{KB}$.
Larger memory regions are needed to create the bucket arrays of the chaining hash tables.
The huge amount of medium sized allocations are requested to materialize tuples using the aforementioned chunks.

Additionally, we measure which operators require the most allocations.
The two main consumer are group by and join operators.
The percentage of allocations per operator for a sequential execution of queries on TPC-DS (SF 100) is shown in the table below:

\vspace{0.25em}
\begin{center}
\begin{tabular}{lccccc}
\toprule
         & Group By & Join   & Set   & Temp  & Other \\ \midrule
By Size  & 61.2\%   & 25.7\% & 4.3\% & 8.4\% & 0.4\% \\
By Count & 77.9\%   & 11.7\% & 8.5\% & 1.8\% & 0.1\% \\ \bottomrule
\end{tabular}
\end{center}
\vspace{0.25em}

To simulate a realistic workload, we use an exponentially distributed workload to determine query arrival times.
We sample from the exponential distribution to calculate the time between two events.
An independent constant average rate $\lambda$ defines the waiting time of the distribution.
In comparison to a uniformly distributed allocation pattern, the number of concurrently active transactions varies.
Thus, a more diverse and complex allocation pattern is created.
The events happen within an expected time interval value of $1/\lambda$ and variance of $1/\lambda^2$.
The executed queries of TPC-DS are uniformly distributed among the start events.
Hence, we are able to test all allocators on the same real-world alike workloads.

Our main-memory query engine allows up to 10 transactions to be active simultaneously.
If more than 10 transactions are queried, the transaction is delayed by the scheduler of our DBMS until the active transaction count is decreased.

%% file: chapters/analysis.tex
\section{Evaluation}
\label{sec:study}

In this section, we evaluate the five allocators on three hardware architectures with different workloads.
We show that the approaches have significant performance and scalability differences.
Additionally, we compare the allocator implementations according to their memory consumption and release strategies which shows memory efficiency and memory fairness to other processes.

We test the allocators on a 4-socket Intel Xeon E7-4870 server (60 cores) with 1 TB of main memory, an AMD Threadripper 1950X (16 cores) with 64 GB main memory (32 GB connected to each die region), and a single-die Intel Core i9-7900X (10 cores) server with 128 GB main memory.
All three systems support 2-way hyperthreading.
These three different architectures are used to analyze the behavior in terms of the allocators' ability to scale on complex multi-socket NUMA systems.

This section begins with a detailed analysis of a realistic workload on the 4-socket server.
We continue our evaluation by scheduling a reduced and increased number of transactions to test the allocators' performance in varying stress scenarios.
An experimental analysis on the different architectures gives insights on the scalability of the five malloc implementations.
An evaluation of the memory consumption and the memory fairness to other processes concludes this section.

\subsection{Memory Consumption and Query Latency} %

\begin{figure}[tb]
  \centering
  \includegraphics{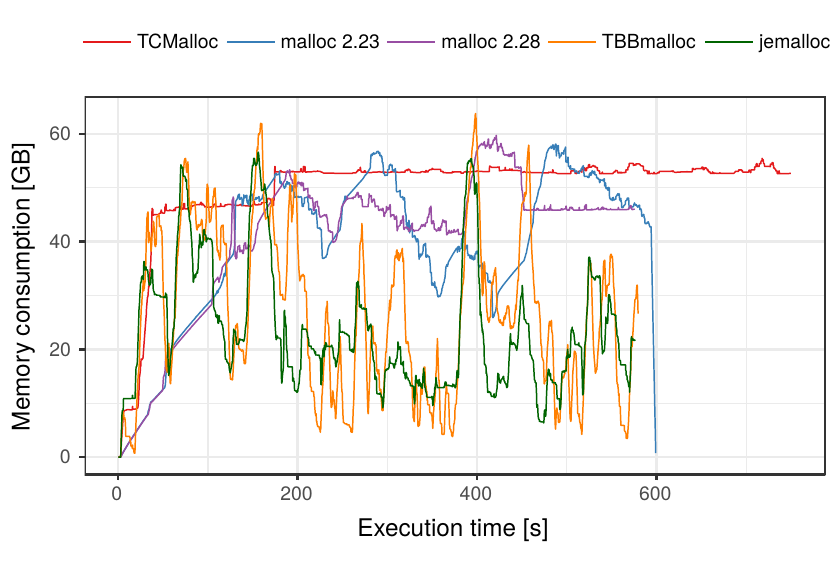}
  \vspace{-2em}
  \caption{Memory consumption over time (4-socket Xeon, $\lambda = 1.25$ q/s, SF 100).}
  \label{fig:mem_10_100p}
\end{figure}

The first experiment measures an exponentially distributed workload to simulate a realistic query arrival pattern on the 4-socket Intel Xeon server.
Figure~\ref{fig:mem_10_100p} shows the memory consumption over time for TCP-DS (SF 100) and a constant query arrival rate of $\lambda = 1.25$ q/s.
Although the same workload is executed, very different memory consumption patterns are measured.
\texttt{TBBmalloc} and \texttt{jemalloc} release most of their memory after query execution.
Both \texttt{malloc} implementations hold a minimum level of memory which increases over time.
\texttt{TCMalloc} releases its memory accurately with \texttt{MADV\_FREE} which is not visible by tracking the system provided resident memory of the database process.
Due to huge performance degradations for tracking the lazy freeing of memory, we show the described release behavior of \texttt{TCMalloc} in Section~\ref{sec:os} separately.
However, the overall performance is reduced due to an increased number of kernel calls.

\begin{figure}
  \centering
  \includegraphics{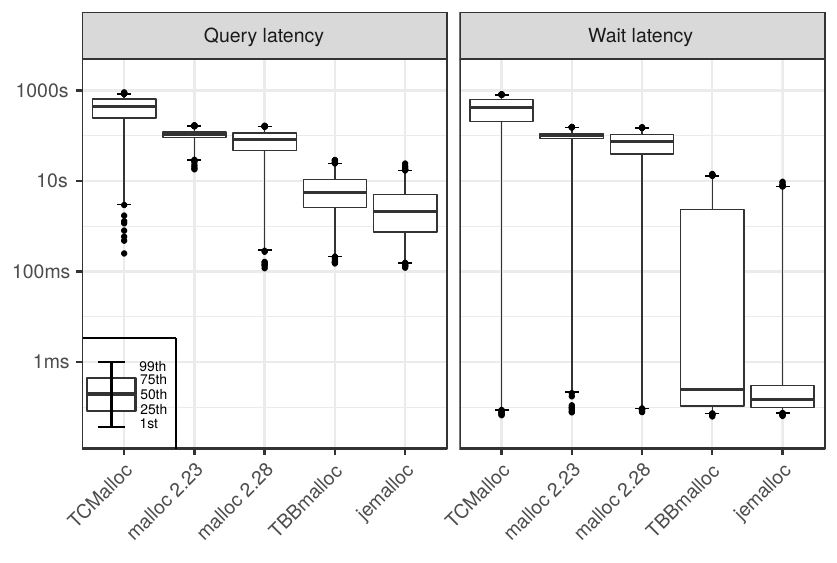}
  \vspace{-2em}
  \caption{Total query latency and wait time (4-socket Xeon, $\lambda = 1.25$ q/s, SF 100).}
  \label{fig:latency}
\end{figure}

For an in-depth performance analysis, the query and wait latencies of the individual queries are visualized in Figure~\ref{fig:latency}.
Although the overall runtime is similar between different allocators, the individual query statistics show that only \texttt{jemalloc} has minor wait latencies.
\texttt{TBBmalloc} and \texttt{jemalloc} are mostly bound by the actual execution of the query.
On the contrary, both glibc \texttt{malloc} implementations and \texttt{TCMalloc} are dominated by the wait latencies.
Thus, the later allocators cannot process the queries fast enough to prevent query congestion.
Query congestion results from the bound number (10) of concurrently scheduled transactions that our scheduler allows to be executed simultaneously.

\begin{table}[]
\centering
\begin{tabular}{l|rrr|r}
\toprule
Allocator   & Local      & Remote      & Total & Page Fault \\ \midrule
malloc 2.28 & 63B,\,100\% & 172B,\,100\% & 236B,\,100\%& 41M, 100\%  \\
jemalloc    & 120\%      & 97\%        & 103\%      & 400\%       \\
TBBmalloc   & 121\%      & 97\%        & 103\%      & 516\%       \\
TCMalloc    & 106\%      & 105\%       & 104\%      & 153\%       \\
malloc 2.23 & 103\%      & 100\%       & 101\%      & 139\%       \\ \bottomrule
\end{tabular}
\caption{NUMA-local and NUMA-remote DRAM accesses and OS page faults (4-socket Xeon, $\lambda = 1.25$ q/s, SF 100).}
\label{tab:normalnumaperformance}
\vspace{-2em}
\end{table}

Because of these huge performance differences, we measure NUMA relevant properties to highlight advantages and disadvantages of the algorithms.
Table~\ref{tab:normalnumaperformance} shows page faults, local and remote DRAM accesses.
All measurements are normalized to the current standard glibc \texttt{malloc 2.28} implementation for an easier comparison.
The two fastest allocators have more local DRAM accesses and significantly more page faults, but have a reduced number of remote accesses.
Note that the system requires more remote DRAM accesses due to NUMA-interleaved memory allocations of the TPC-DS base relations.
Thus, the highly increased number of local accesses change the overall number of accesses only slightly.
Minor page faults are not crucially critical since both \texttt{jemalloc} and \texttt{TBBmalloc} release and acquire their pages frequently.
Consequently, remote accesses for query processing are the major performance indicator.
Because \texttt{TCMalloc} reuses \texttt{MADV\_FREE} pages, the number of minor page faults remains small.

\begin{figure}[tb]
  \centering
  \includegraphics{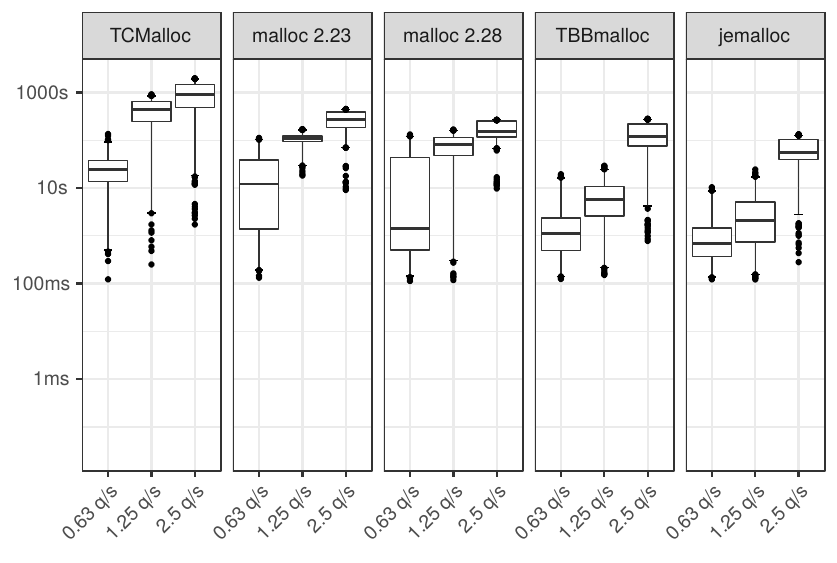}
  \vspace{-2em}
  \caption{Query latency distributions for different query rates (4-socket Xeon, SF 100).}
  \label{fig:latencyhighlow}
  \vspace{-2em}
\end{figure}

\subsection{Performance with Varying Stress Levels}

In the previous workload, only two allocators were able to efficiently handle the incoming queries.
This section evaluates the effects for a varying constant rate $\lambda$.
We analyze two additional workloads that use the rates $\lambda = 0.63$ and $\lambda = 2.5$ queries per second.
Thus, we respectively increase and decrease the average waiting time before a new query is scheduled by a factor of 2.

Figure~\ref{fig:latencyhighlow} shows the query latencies of the three workloads.
The results for the reduced and increased waiting times confirm the previous observations.
The allocators have the same respective latency order in all three experiments.
\texttt{jemalloc} performs best again for all workloads, followed by \texttt{TBBmalloc}.

All query latencies are dominated by the wait latencies in the $\lambda = 2.5$ workload due to frequent congestions.
With an increased waiting time ($\lambda = 0.63$) between queries, the glibc \texttt{malloc 2.28} implementation is able to reduce the median latency to a similar level as \texttt{TBBmalloc}.
However, the query latencies within the third quantile vary vastly.
\texttt{TCMalloc} and \texttt{malloc 2.23} are still not able to process the queries without introducing long waiting periods.

\subsection{Scalability}
After analyzing the allocators' perfromance on the 4-socket Intel Xeon architecture, this section focuses on the scalability of the five dynamic memory allocators.
Therefore, we execute an exponentially distributed workload with TPC-DS (SF 10) on the NUMA-scale 60 core Intel Xeon server, the 16 core AMD Threadripper (two die regions), and the single-socket 10 core Intel Skylake X.

\begin{figure}[tb]
  \centering
  \includegraphics{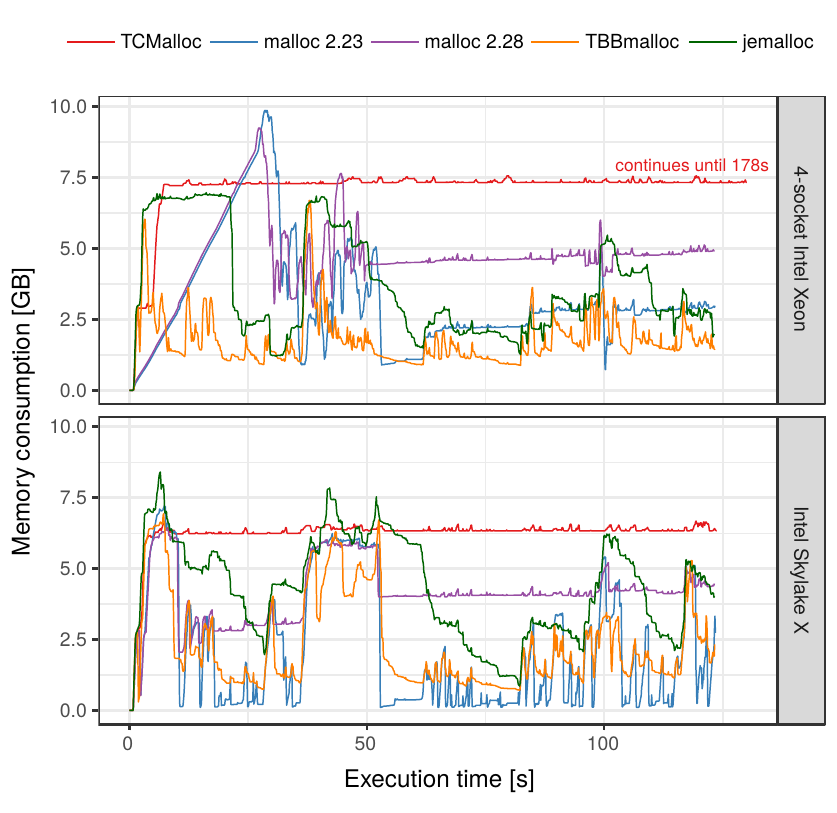}
  \vspace{-2em}
  \caption{Memory consumption over time ($\lambda = 6$ q/s, SF 10).}
  \label{fig:sxtr_mem_10_10}
  \vspace{-1em}
\end{figure}

Figure~\ref{fig:sxtr_mem_10_10} shows the memory consumption during the workload execution.
Since the AMD Threadripper has a very similar memory consumption pattern to the Intel Skylake X, we only show the 4-socket Intel Xeon and the single-socket Intel Skylake.
Most notable are the differences of both glibc \texttt{malloc} implementations.
These two allocators have a very long initialization phase on the 4-socket system, but are able to allocate their initial memory as fast as the other ones on the single-socket system.
Due to more cores and the resulting different access pattern, the decay-based deallocation pattern of \texttt{jemalloc} differs slightly in the beginning.
However, \texttt{jemalloc}'s decay-based purging reduces the memory consumption on both architectures considerably.
\texttt{TCMalloc} cannot process all queries in the same time frame as the other allocators on the 4-socket system whereas it finishes at the same time on Skylake.

\begin{figure}[tb]
  \centering
  \includegraphics{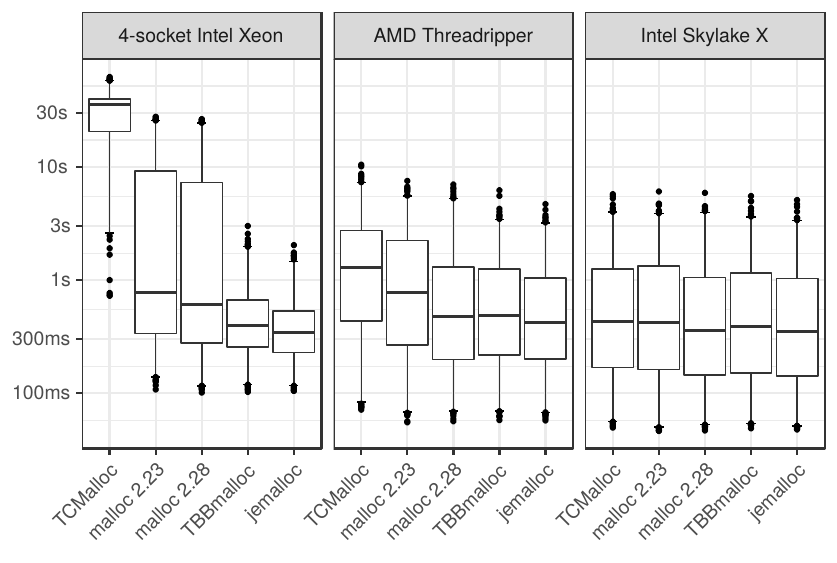}
  \vspace{-2em}
  \caption{Query latencies ($\lambda = 6$ q/s, SF 10).}
  \label{fig:reallat_med_10_10}
  \vspace{-0.25em}
\end{figure}

Especially the query latencies differ vastly between the architectures.
In Figure~\ref{fig:reallat_med_10_10}, we show the latencies for the $\lambda = 6$ q/s workload.
The more cores are utilized, the larger are the latency differences between the allocators.
On the single-socket Skylake X, all the allocators have very similar performance.
Besides having more cores, AMD's Threadripper uses two memory regions which requires a more advanced placement strategy to obtain fast accesses.
In particular, \texttt{TCMalloc} and \texttt{malloc 2.23} without a thread-local cache have a reduced performance.
The latency variances are reduced on the Threadripper but the overall latencies are worse in comparison to the Skylake architecture.

Yet, the most interesting behavior is introduced by the multi-socket Intel Xeon.
It has both the best and worst overall query performance.
\texttt{jemalloc} and \texttt{TBBmalloc} execute the queries with the overall lowest latencies and smallest variance.
On the other hand, \texttt{TCMalloc} is worse by more than 10x in comparison to any other allocator.
Both glibc implementations have a similar median performance but incur high variance such that a reliable query time prediction is impossible.

The experiments show that both \texttt{jemalloc} and \texttt{TBBmalloc} are able to scale to large systems with many cores.
\texttt{TCMalloc}, on the other hand, has significant performance loss on larger servers.

To validate our findings, we evaluate a subset of the queries on MonetDB 11.31.13~\cite{DBLP:journals/debu/IdreosGNMMK12}.
We observe a performance boost by using \texttt{jemalloc} on MonetDB; however, the differences are smaller because our DBMS parallelizes better and thus utilizes more cores.

\subsection{Memory Fairness}
\label{sec:os}

\begin{table}[t]
\begin{tabular}{l | cc | cc}
\toprule
            & \multicolumn{2}{l|}{peak total} & \multicolumn{2}{l}{average total} \\
Allocator   & requested & measured\footnotemark & requested & measured \\ \midrule
TCMalloc    & 55.7 GB & 58.1 GB & 17.8 GB & 53.7 GB \\
malloc 2.23 & 61.4 GB & 61.0 GB & 26.2 GB & 41.3 GB \\
malloc 2.28 & 61.5 GB & 62.6 GB & 20.2 GB & 42.5 GB \\
TBBmalloc   & 55.7 GB & 55.7 GB & 15.9 GB & 27.9 GB \\
jemalloc    & 58.6 GB & 59.4 GB & 11.1 GB & 24.7 GB \\ \bottomrule
\end{tabular}
\caption{Memory usage (4-socket Xeon, $\lambda = 1.25$ q/s, SF 100).}
\label{tab:memory100_10}
\vspace{-2.5em}
\end{table}

Many DBMS run alongside other processes on a single server.
Therefore, it is necessary that the query engines are fair to other processes.
In particular, the memory consumption and the memory release pattern are good indicators of the allocators' memory fairness.

\begin{figure}[tb]
  \centering
  \includegraphics{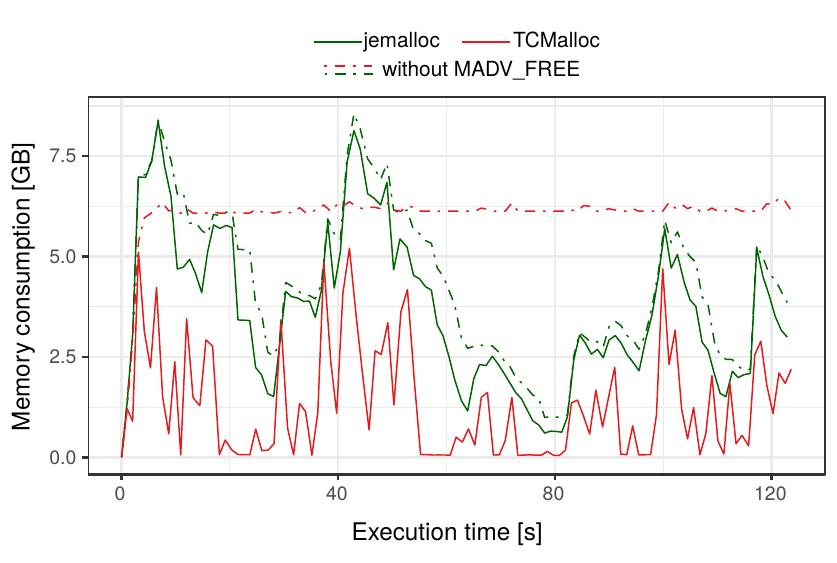}
  \vspace{-2.5em}
  \caption{Memory consumption over time with subtracted \texttt{MADV\_FREE} pages ($\lambda = 6$ q/s, SF 10).}
  \label{fig:sx_lazy_mem_10_10}
  \vspace{-1em}
\end{figure}

Our DBMS is able to track the allocated memory regions with almost no overhead.
Hence, we can compare the measured process memory consumption with the requested one.
The used memory differs between the allocators due to the performance and scalability properties although we execute the same set of queries.
Table~\ref{tab:memory100_10} shows the peak and average memory consumption for the $\lambda = 1.25$ q/s workload (SF 100) on the 4-socket Intel Xeon.
\footnotetext{Due to chunk-wise allocation with unfaulted pages and measurement delays the measured amount of memory can be slightly smaller than the requested one.}
The peak memory consumption is similar for all tested allocators.
On the contrary, the average consumption is highly dependant on the used allocator.
Both glibc \texttt{malloc} implementations demand a large amount of average memory.
\texttt{jemalloc} requires less average memory than \texttt{TBBmalloc}.
However, the DBMS requested average memory is also higher for the allocators with increased memory usage.
Although the consumption of \texttt{TCMalloc} seems to be higher, it actually uses less memory than the other allocators.
This results from the direct memory release with \texttt{MADV\_FREE}.
The tracking of \texttt{MADV\_FREE} calls on the 4-socket Intel Xeon is very expensive and would introduce many anomalies for both performance and memory consumption.
Therefore, we analyze the madvise behavior on the single-socket Skylake X that is only affected slightly by the \texttt{MADV\_FREE} tracking.
The memory consumption with the $\lambda = 6$ q/s workload (SF 10) is shown in Figure~\ref{fig:sx_lazy_mem_10_10}.
The only two allocators that use \texttt{MADV\_FREE} to release memory are \texttt{jemalloc} and \texttt{TCMalloc}.
The measured average memory curve of \texttt{TCMalloc} follows the DBMS required curve almost perfectly.
\texttt{jemalloc} has a 15\% reduced consumption if the \texttt{MADV\_FREE} pages are subtracted from the memory consumption.

%% file: chapters/conclusion.tex
\section{Conclusions}
\label{sec:summary}

In this work, we provided a thorough experimental analysis and discussion on the impact of dynamic memory allocators for high-performance query processing.
We highlighted the strength and weaknesses of the different state-of-the-art allocators according to scalability, performance, memory efficiency, and fairness to other processes.
For our allocation pattern, which is probably not unlike to that of most high-performance query engines, we can summarize our findings as follows:

\vspace{0.15em}
\begin{center}
\begin{tabular}{lcccc}
\toprule
            & scalable & fast   & mem.\,fair & mem.\,efficient \\ \midrule
TCMalloc    & $--$     & $\sim$ & $++$ &    +     \\
malloc 2.23 & $-$      & $\sim$ & $+$  &  $\sim$  \\
malloc 2.28 & $\sim$   & $+$    & $-$  &  $\sim$  \\
TBBmalloc   & $+$      & $\sim$ & $++$ &    +     \\
jemalloc    & $++$     & $+$    & $+$  &    +     \\ \bottomrule
\end{tabular}
\end{center}
\vspace{0.15em}

As a result of this work, we use \texttt{jemalloc} as the standard allocator for our DBMS.

\vspace{1em}
This  project  has  received  funding  from  the  European  Research Council (ERC) under the European Union’s Horizon 2020 research and innovation programme (grant agreement No 725286). \hfill\includegraphics[height=3mm]{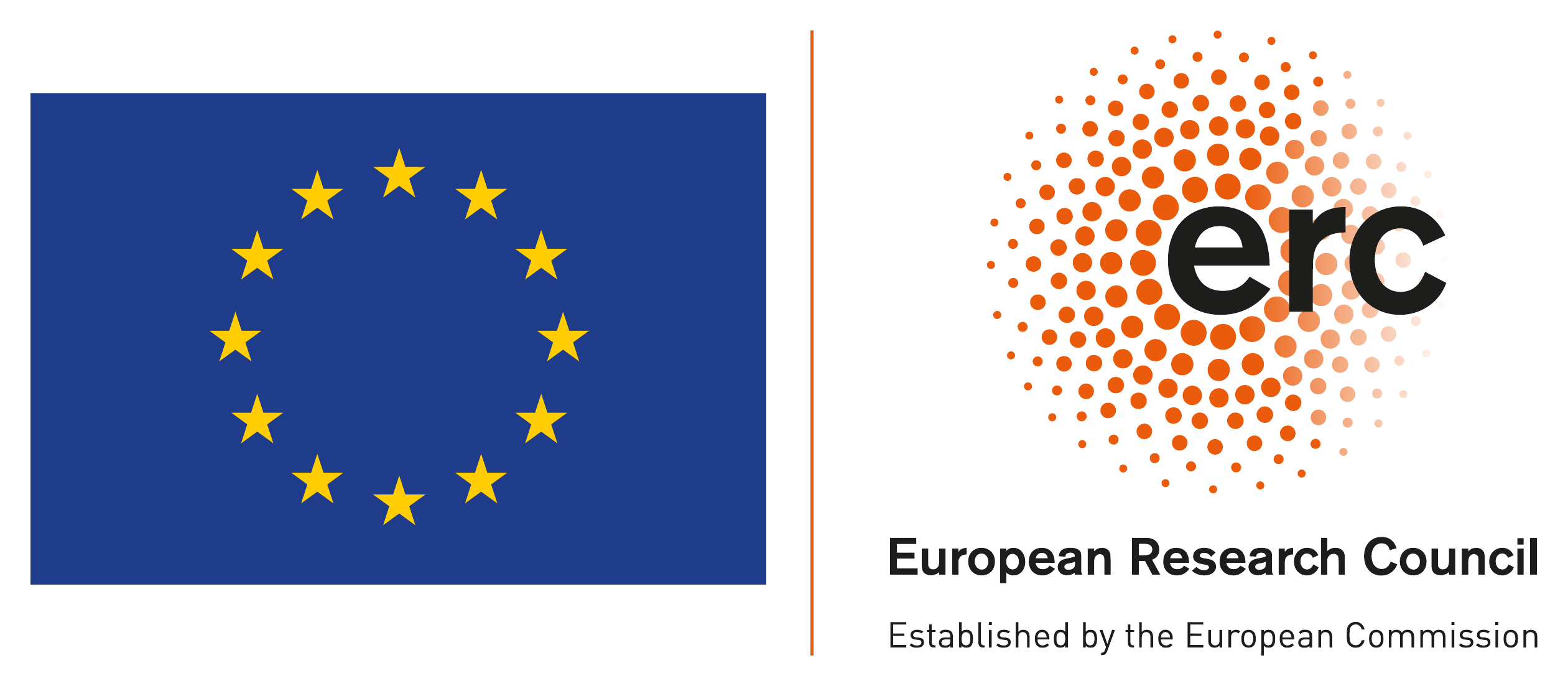}